\documentclass[
reprint,
superscriptaddress,
nofootinbib,
amsmath,amssymb,
aps,
pra,
twocolumn,
floatfix,
]{revtex4-2}

\usepackage[utf8]{inputenc}
\usepackage[english]{babel}

\usepackage{amsmath}
\usepackage{amssymb}
\usepackage{bbm}
\usepackage{bm}
\usepackage{braket}
\usepackage{graphicx}
\usepackage{dsfont}

\usepackage[usenames,dvipsnames]{xcolor}
\usepackage[colorlinks=true,citecolor=MidnightBlue,linkcolor=MidnightBlue,urlcolor=MidnightBlue]{hyperref}

\usepackage[capitalize]{cleveref}
\begin{document}

\title{Hardware-Efficient Preparation of Graph States on Near-Term Quantum Computers}

\author{Sebastian Brandhofer}
\affiliation{Institute of Computer Architecture and Computer Engineering, University of Stuttgart, 70569 Stuttgart, Germany}
\affiliation{Center for Integrated Quantum Science and Technology (IQST), University of Stuttgart, 70569 Stuttgart, Germany}
\author{Ilia Polian}
\affiliation{Institute of Computer Architecture and Computer Engineering, University of Stuttgart, 70569 Stuttgart, Germany}
\affiliation{Center for Integrated Quantum Science and Technology (IQST), University of Stuttgart, 70569 Stuttgart, Germany}
\author{Stefanie Barz}
\affiliation{Center for Integrated Quantum Science and Technology (IQST), University of Stuttgart, 70569 Stuttgart, Germany}
\affiliation{Institute for Functional Matter and Quantum Technologies, University of Stuttgart, 70569 Stuttgart, Germany}
\author{Daniel Bhatti}
\affiliation{Center for Integrated Quantum Science and Technology (IQST), University of Stuttgart, 70569 Stuttgart, Germany}
\affiliation{Institute for Functional Matter and Quantum Technologies, University of Stuttgart, 70569 Stuttgart, Germany}
\affiliation{Networked Quantum Devices Unit, Okinawa Institute of Science and Technology Graduate University, Okinawa, Japan}

\begin{abstract}
Highly entangled quantum states are an ingredient in numerous applications in quantum computing.
However, preparing these highly entangled quantum states on currently available quantum computers at high fidelity is limited by ubiquitous errors.
Besides improving the underlying technology of a quantum computer, the scale and fidelity of these entangled states in near-term quantum computers can be improved by specialized compilation methods.
In this work, the compilation of quantum circuits for the preparation of highly entangled architecture-specific graph states is addressed by defining and solving a formal model.
Our model incorporates information about gate cancellations, gate commutations, and accurate gate timing to determine an optimized graph state preparation circuit.
Up to now, these aspects have only been considered independently of each other, typically applied to arbitrary quantum circuits.
We quantify the quality of a generated state by performing stabilizer measurements and determining its fidelity.
We show that our new method reduces the error when preparing a seven-qubit graph state by 3.5x on average compared to the state-of-the-art Qiskit solution.
For a linear eight-qubit graph state, the error is reduced by 6.4x on average.
The presented results highlight the ability of our approach to prepare higher fidelity or larger-scale graph states on gate-based quantum computing hardware.
\end{abstract}

\maketitle

\section{Introduction}

Highly entangled multi-qubit graph states, e.g., linear graph states or 2D cluster states, are essential for a large number of quantum applications \cite{Hein2006,Liao2022,Pathumsoot2020,Ruckle2023,Raussendorf2001,Raussendorf2003,qaoa,vqe,Ferguson2021,Chan2024}.
Not only are they being used in quantum error correction \cite{Hein2006,Liao2022} or quantum communication protocols \cite{Pathumsoot2020,Ruckle2023}.
They also build the basis for various quantum computing techniques such as one-way quantum computing \cite{Raussendorf2001,Raussendorf2003}, variational quantum algorithms \cite{qaoa,vqe}, and, as recently shown, combinations of the two \cite{Ferguson2021,Chan2024}.

Due to their multi-particle entanglement, graph states are used to demonstrate non-classical behavior
and thus often employed to benchmark quantum computers~\cite{Wang2018,Mooney2019,Mooney2021,Mooney2021GHZ,Mackeprang2023}.
A special group of graph states are \textit{native graph states}. Native graph states are graph states where the mathematical graph is in accordance with the qubit connectivity of the respective quantum machine~\cite{Mooney2021}.
They can serve as efficient and scalable entanglement benchmarks and have led to the demonstration of full bipartite entanglement using up to 65 qubits~\cite{Wang2018,Mooney2019,Mooney2021}.
The preparation of hardware-efficient and high-fidelity native graph states on near-term gate-based quantum computers is the focus of this work.

Besides benchmarking, one can use native graph states for efficiently preparing non-native graph states. For example, by employing local complementation, i.e., local Clifford operations on the quantum state, non-native graph states can be generated using the same set of qubits as the initial state~\cite{Hein2006}.
Moreover, depending on the specific form of the target state, measuring out particular qubits of a larger (native) graph state, can increase the overall preparation efficiency and enlarge the range of accessible graph states~\cite{Meignant2019,Pathumsoot2020,deJong2024}.

Current quantum computers suffer from relatively large and heterogeneous errors that limit the ability to prepare high-fidelity graph states~\cite{Preskill2018, arsonisq}.
Error mitigation and quantum circuit compilation can reduce errors of state preparation and improve the fidelity of graph state applications on near-term noisy, intermediate-scale quantum (NISQ) computers~\cite{bravyi_readout_error_mitigation, olsq, ibmqx_efficient,Preskill2018}.

In this work, we present a compilation method for improving the quantum circuits used to prepare hardware-efficient graph states on near-term IBM quantum computers.
We define a formal model that is based on the structure of a given type of graph state.
This model allows us to:
\begin{itemize}
    \item prepare hardware-efficient native graph states using a minimal duration of or number of single-qubit gates
    %\item considers the present error characterization and basis gates of an IBM quantum computer.
    \item use gate commutation to enable a reordering of quantum gates for the minimization of quantum circuit duration
    \item employ gate cancellation to identify and omit excess single-qubit gates
    \item consider accurate timing information for exact minimization of the graph state circuit duration beyond the minimization of quantum circuit depth
\end{itemize}
Up to now, these aspects have only been considered independently of each other and are typically applied to arbitrary quantum circuits~\cite{sat_commute, olsq_commute, toqm}.
By combining the different aspects and applying them to a special class of quantum circuits, i.e., graph state generation circuits, we present an approach that allows us to scale to larger-scale problem instances while expected to yield higher-fidelity graph state preparation circuits.
The resulting quantum compilation method is evaluated for graph state preparations on the IBM quantum computer \texttt{ibmq\_ehningen}~\cite{ehningen}.
While the evaluation in this work is restricted to graph states and IBM quantum computers, the presented approach can be generalized to other entangled states and NISQ computers.

Please note that this work builds on one of the winning contributions to the 2020 IBM Quantum Open Science Prize~\cite{Challenge}.

\section{Near-Term Quantum Computers\label{sec nisq}}

In theory, an $n$-qubit quantum computer can arbitrarily prepare, manipulate, and measure an $n$-qubit state given by 
\begin{equation}
    \ket{\psi} = \sum_{i} \alpha_{i} \ket{i},
\end{equation}
with $\sum_{i} |\alpha_{i}|^2 = 1$.
In reality, however, near-term quantum computers are often characterized by a limited number of qubits in the range of a couple of dozens to a couple of hundreds, a restricted qubit connectivity, denoted by its topology (see \cref{fig graph state}), and a heterogeneous qubit quality with relatively short coherence times~\cite{Preskill2018}.

Decoherence, imperfect quantum gates, and measurements lead to errors during the computation of a quantum algorithm~\cite{here}.
In addition, qubits typically exhibit a heterogeneous quality, leading to varying coherence times, quantum gate durations, and errors on the same device and time step \cite{not-all-qubits-are-equal}.
As an example, in the 27-qubit near-term IBM quantum computer used in the evaluation section of this paper, the coherence time of the qubits varies
from 0.4~$\mu$s to 343~$\mu$s per qubit, the two-qubit quantum gate error rate varies
from 0.4\% to 21\%, and the two-qubit gate duration varies
from 181~ns to 587~ns.
It is, therefore, crucial to not only consider the logical depth of a quantum circuit, i.e., the length of its critical path from the inputs to the outputs \cite{olsq, ibmqx_efficient}, when minimizing the duration of a quantum computation, but also consider the accurate timing of the quantum gates in the quantum circuit.

\section{Graph States\label{sec graph_states}}

\begin{figure}[t!]
    \centering
    \includegraphics[width=1.0\linewidth]{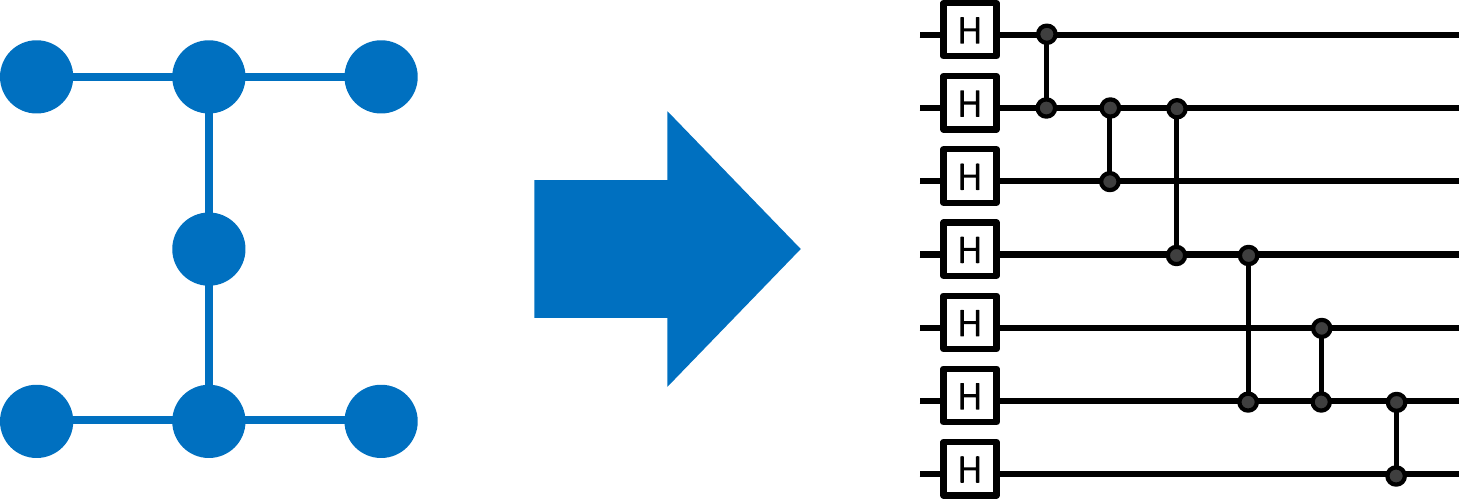}    
    \caption{
    Left: Seven-qubit graph structure. A graph state consists of vertices (= qubits) and edges (= two-qubit entanglement). In this work, we generate graph states identical to the quantum computer's topology graph. In the shown example, qubits are arranged in a rotated "H" such that two qubits have three neighbors, one qubit has two neighbors and four qubits have one neighbor. Neighboring qubits can interact with each other directly. Right: Corresponding preparation circuit of a seven-qubit graph state. To generate graph states, we use Hadamard (H) gates preparing each qubit in the state $\ket{+}$, followed by two-qubit CZ gates along the edges of the graph generating the entanglement.
    }
    \label{fig graph state}
\end{figure}

A graph state $\ket{\mathcal{G}}$ is a multi-qubit entangled quantum state, which can be described by a mathematical graph $\mathcal{G}=(V,E)$ (see left part of \cref{fig graph state} and, e.g.,~\cite{Hein2006}).
The graph consists of $n$ vertices in the vertex set $V$, which correspond to qubits $a_1,\ldots,a_n$,
and a set of edges $E$,
which indicate entanglement between the connected qubits $(a_i,a_j)\in E$ . 

Mathematically, graph states can be described using the so-called stabilizer formalism. This formalism defines one stabilizer operator $S_{a_i}$ per vertex $a_i \in V$~\cite{Hein2006}:
\begin{equation}
\label{eq:stabilizer}
    S_{a_i} = X_{a_i} \prod_{a_j \in N_{a_i}} Z_{a_j} ,
\end{equation}
where $N_{a_i}$ describes the vertices adjacent to $a_i$, and $X_{a_i}$ ($Z_{a_j}$) denotes the Pauli $X$ ($Z$) operator on the $i$th ($j$th) qubit.

Using \cref{eq:stabilizer}, one can now define the graph state $\ket{\mathcal{G}}$ as the unique eigenstate of all $S_{a_i}$ with eigenvalue $+1$, i.e.,
\begin{align}
    S_{a_i} \ket{\mathcal{G}} = +\ket{\mathcal{G}}, \ \forall i .
\end{align}
Measuring the stabilizer elements individually,
therefore, allows for determining the graph state fidelity using only $2^n$ measurements instead of $3^n$ measurements required for a full quantum state tomography~\cite{Kiesel2005}.

Generating graph states can be accomplished using Hadamard (H) gates and controlled Pauli Z (CZ) gates (see right part of \cref{fig graph state} and, e.g.,~\cite{Mooney2021}).
First, each qubit, initially in the state $\ket{0}$, is prepared in the state $\ket{+}= \text{H} \ket{0} = 1/\sqrt{2}(\ket{0}+\ket{1})$. Then the 
entanglement is realized by two-qubit entanglement operations, i.e., CZ gates along the edges $E$. This allows one to write every graph state in the form
\begin{equation}
    \ket{\mathcal{G}} = \prod_{(a_i,a_j)\in E} \text{CZ}_{(a_i,a_j)} \ket{+}\ket{+}\dots \ket{+} .
\end{equation}
In this work, we focus on \textit{native graph states} and \textit{linear graph states}.
Native graph states have a \textit{graph structure} $\mathcal{G}$ that is (subgraph) isomorphic \cite{subgraph_isomorphism} to the topology graph of the respective quantum device. This means that the edge set $E$ in $\mathcal{G}$ can be mapped to a connected subset of the edges in the topology graph \cite{Mooney2021}.
Linear graph states have a graph structure that corresponds to a path graph, i.e., two vertices have degree one while the remaining vertices have degree two.
An elegant way of preparing non-native graph states is to use local complementation~\cite{Hein2006}.

\section{Compilation for Graph State Preparation\label{sec method}}

Our compilation method is based on a formal model that considers the graph state structure, topology, and error characterization of a given quantum computer to yield an improved graph state preparation circuit. The improved graph state preparation circuit is compiled while considering gate commutation relations, gate cancellations, and accurate timing information provided by the quantum computing operator.

\begin{figure}[b]
    \centering
    \includegraphics[width=1.0\linewidth]{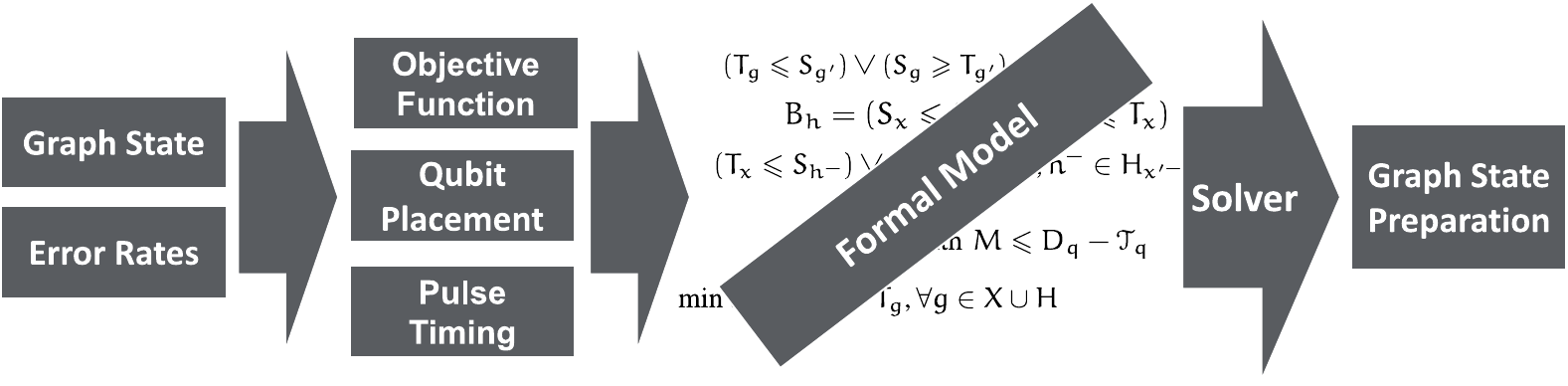}       
    
    \caption{Individual steps of the developed graph state preparation compilation method.}
    \label{fig flow chart}
\end{figure}

\cref{fig flow chart} depicts the individual steps of the developed quantum circuit compilation method for graph state preparation.
First, the structure of a graph state and the current error characterization are used to determine the placement of the graph state qubits onto the physical qubits of the target quantum computer.
This step is realized by methods such as mapomatic \cite{mapomatic} that consider the product of current gate fidelities on the quantum computer to determine an improved placement.

Then, the placement information is used together with the structure of a given graph state, accurate quantum gate timing information, and an objective function to inform the generation of a formal model.
The formal model can then be solved to yield an optimized graph state preparation circuit \cite{z3-solver}.
The formal model consists of variables that represent valid preparation circuits if the value assignments of the model variables satisfy the constraints defined in this model.
The solver then optimizes the valid assignment to the model variables respect to given objective functions.
The resulting assignment is provably optimal, i.e. the solver always finds the global minimum.

The compilation method assumes native graph states as defined in \cref{sec graph_states} and a target quantum computer with a basis gate set that includes CNOT and Hadamard gates.
The CZ quantum gate needed in the construction of graph states is thus represented in this basis gate set by applying Hadamard gates before and after a CNOT gate on the CNOT's target qubit for the remainder of this work \cite{qc10th}.
Notably, the steps in this work focus on the currently available IBM quantum computers, where the CNOT gate is available natively, and the Hadamard gate is available through other single-qubit gates.
The method developed in this work can be adapted for different basis gate sets \cite{qiskit}.

\subsection{Model Variables} \label{sec model variables}

The developed model has the following model variables for a graph state with quantum gates 
$G=F\cup H$, where $F$
is the set of two-qubit quantum gates and $H$ is the set of Hadamard gates in the graph state preparation circuit:
\begin{itemize}
    \item $C$ --- the set of Boolean variables representing the two different directions of CNOT gates. As the only two-qubit quantum gates in graph states are CZ quantum gates originally, the role of the target qubit and the control qubit is exchangeable, i.e., the direction of a CNOT can be set arbitrarily. The direction of CNOT gates has a significant impact on the duration of the CNOT gate \cite{ehningen} and also affects how many quantum gates in the quantum circuit can be canceled.
    \item $S$ --- the set of real variables representing the start times of each quantum gate in the graph state preparation circuit. The set of variables $S_H$ represents the start times of Hadamard gates, and $S_F$ represents the start times of CNOT gates with $S = S_{H} \cup S_{F}$.
    \item $T$ --- the set of real variables representing the end time of each quantum gate in the graph state preparation circuit. As above, the set of variables $T_H$ represents the end times of Hadamard gates, and $T_F$ represents the end times of CNOT gates with $T = T_{H} \cup T_{F}$.
    \item $B$ --- the set of Boolean variables indicating whether a Hadamard gate is canceled due to a directly subsequent or preceding Hadamard gate. In the native graph states considered in this work, there is no pair of two-qubit quantum gates on the same set of qubits. Therefore, only Hadamard gates can cancel out. 
\end{itemize}
In addition, let $H_f$ be the set of Hadamard quantum gates that transform the two-qubit quantum gate $f\in F$ into the CZ quantum gate with $H_{f} = H_{f^{-}} \cup H_{f^{+}}$, where $H_{f^{-}}$ are single-qubit quantum gates occurring before and $H_{f^{+}}$ are single-qubit quantum gates occurring after the computation of the two-qubit quantum gate $f$.
For the IBM quantum computers considered in this work, the sets $H_{f^{-}}$ and $H_{f^{+}}$ consist of only one Hadamard gate each that is applied to the target qubit of the CNOT gate.
The solver software used in this work \cite{z3-solver} can address the real variables in sets $T$ and $S$.
Alternatively, these real-valued variables can be discretized \cite{openpulse}.

\subsection{Model Constraints}

Model constraints guarantee that a satisfying assignment to the model variables yields a valid quantum circuit.
For the native graph state preparation circuits considered in this work, the constraints must assign the accurate duration to gates in the circuit, cancel Hadamard gates in the correct situations, and ensure the correct order of non-commuting quantum gates while allowing an arbitrary order of commuting quantum gates.
First, the duration of the quantum gates in the graph state preparation quantum circuit is modeled depending on the chosen direction and the accurate timing information specified by the quantum computer vendor.

\begin{equation} \label{eq constr a}
T_{g} - S_{g} = \left(d_{g} \wedge C_{g}\right) \vee \left(\Bar{d_{g}} \wedge \neg C_{g} \right),
\end{equation}
for a two-qubit quantum gate $g$ that has a gate duration of $d_g$ in the direction $C_g$ and $\Bar{d_g}$ in the other direction.
The equations for a single-qubit quantum gate $g'$ are similar but do not have a direction and, as such, are directly assigned their gate duration $d_{g'}$ according to the difference of $T_{g'}$ and $S_{g'}$.

Next, the temporal order of quantum gates needs to be considered in a valid graph state preparation quantum circuit.
In general, CZ quantum gates commute with each other, so the temporal order of these quantum gates can be set arbitrarily as long as a qubit is not participating in two CZ quantum gates at once \cite{qc10th}.
Thus, as the CZ gate is represented by a CNOT gate and two Hadamard gates on the target qubit, the set of CNOT and corresponding Hadamard gates also commute with each other.
Furthermore, two subsequent Hadamard gates cancel each other out.

The following equations capture the exact timings of Hadamard gates and CNOT gates.
First, the timing of the CNOT gate $f$ with the sandwiched Hadamard gates is fixed by:
\begin{equation}\label{eq constr b}
    T_{h} \leq S_{f}, \forall h\in H_{f^{-}},
\end{equation}
and
\begin{equation}\label{eq constr c}
    S_{h} \geq T_{f}, \forall h\in H_{f^{+}}.
\end{equation}
Note that the temporal order of quantum gates inside the sets $H_i$ does not need to be fixed in our case because they consist of only one gate.
Furthermore, we distinguish two cases of two quantum gates $f$ and $f'$ overlapping on the same qubit.
First, if the control qubit or the target qubit of the quantum gates $f$ and $f'$ overlap, then
\begin{equation}\label{eq constr d}
    \left( T_{f} \leq S_{f'}\right) \vee \left( S_{f} \geq T_{f'}\right)
\end{equation}
must hold, i.e., gate $f$ must end before gate $f'$ or gate $f$ must start after gate $f'$.
In this case, the Hadamard gate on the target qubit of the CNOT gate cancels with the Hadamard gate $h$.
Thus, the single-qubit quantum gates $H_f$ associated with a two-qubit quantum gate $f$ can overlap temporally with the computation of a different two-qubit quantum gate $f'$ or its single-qubit quantum gates $H_{f'}$.
A conflicting temporal assignment of quantum gates, i.e., one qubit would need to participate in multiple quantum gates at once, can occur for gates that cancel.
This is reflected through variables $B$ that indicate which quantum gates are canceled such that the assigned computation time is irrelevant.

Likewise, the following set of equations is enforced if the control qubit of quantum gate $f$ overlaps with the target qubit of quantum gate $f'$:
\begin{equation}\label{eq constr e}
    %\left(T_{h} \leq S_{f'} \right)  \vee \left( S_{h'}\geq T_{f'}\right), h\in H_{f^{+}}, h' \in H_{f^{-}}
    \left(T_{f} \leq S_{h^{-}} \right)  \vee \left( S_{f} \geq T_{h^{+}}\right), h^{-}\in  H_{f'^{-}}, h^{+} \in H_{f'^{+}}.
\end{equation}
As Hadamard gates do not commute with CNOT gates, these equations limit the temporal placement of Hadamard gates to a preceding or a succeeding CNOT gate $f'$ that is overlapping with gate $f$.
In addition, the computation time of the CNOT gate cannot overlap with the computation time of a Hadamard gate.

The cancellation of Hadamard gates can be expressed by
\begin{equation}\label{eq constr f}
   B_{h} = \left( S_{f} \leq S_{h} \right) \wedge \left( T_{h} \leq T_{f} \right),
\end{equation}
i.e., a Hadamard gate $h$ is canceled if its computation time overlaps with a two-qubit quantum gate $f$ whose target qubit occurs on the same qubit as the Hadamard gate $h$.
Further domain constraints, e.g. restricting the start and end times to non-negative reals, are omitted.

\subsection{Objective Functions}

We develop four objective functions for the compilation of hardware-efficient graph state preparation quantum circuits.
A first objective to achieve is to minimize the number of Hadamard gates along with the required CNOT gates for the preparation of a given native graph state.
This is realized by maximizing the number of gate cancellations:
\begin{equation} \label{eq cancellation}
    \max \sum B_{i}.    
\end{equation}

In addition, reducing the effect of decoherence can be achieved by minimizing the overall duration of the graph state preparation circuit \cite{toqm}.
The duration of a quantum circuit can be minimized through
\begin{equation} \label{eq runtime}
    \min \mathcal{T} \text{ with } \mathcal{T} \geq T_{g}, \forall g\in F\cup H , 
\end{equation}
where $T_{g}$ is the individual end time of the quantum gate $g$ and $\mathcal{T}$ is an auxiliary variable.

An additional figure of merit for the reduction of the impact of decoherence is the maximization of the 'remaining' coherence time on a qubit as given by the difference between the determined qubit coherence time and the circuit duration on each qubit.
The remaining coherence time can be expressed by
\begin{equation} \label{eq decoherence}
    \max M, \text{ with } M \leq D_{q} - \mathcal{T}_{q},
\end{equation}
where $D_{q}$ is the coherence time on qubit $q$ and $\mathcal{T}_q$ is the end time of the last quantum gate on qubit $q$. The variable $\mathcal{T}_q$ can be determined analogously to \cref{eq runtime}.

The objective function concerns the reduction of crosstalk errors that occur when a two-qubit quantum gate is performed on a pair of qubits where neighboring qubits are not idle~\cite{crosstalk-mapping, crosstalk-ketterer, crosstalk_rudinger}.
These types of context-dependent errors can have a large impact on the fidelity of a quantum state preparation and thus pose a potential for minimization by
\begin{equation} \label{eq crosstalk}
    \left(T_{g} \leq S_{g'}\right) \vee \left(S_{g} \geq T_{g'}\right)
\end{equation}
for all quantum gates $g$ and $g'$ where the quantum gate $g'$ acts on neighboring qubits.
These can be determined, e.g., by the methods introduced in \cite{crosstalk-ketterer}.

\subsection{Deriving a Quantum Circuit From a Solved Formal Model}

A graph state preparation quantum circuit is exactly determined by the variables introduced in \cref{sec model variables}.
The time and qubits of a quantum gate in the graph state preparation circuit are exactly fixed by the developed formal model, i.e., a quantum circuit representation can be derived in linear time by inspecting the model variable assignments.

A satisfiability modulo theories (SMT) solver such as the Z3 SMT solver can determine an assignment to these model variables that satisfies the constraints specified in \cref{eq constr a,eq constr b,eq constr c,eq constr d,eq constr e,eq constr f} and is optimal with respect to a specified objective function such as defined in \cref{eq cancellation,eq runtime,eq decoherence,eq crosstalk} \cite{z3-solver}.

\section{Evaluation\label{sec eval}}

In this section, we evaluate our compilation method on a seven-qubit native graph state (see \cref{fig graph state}) and on linear graph states with three to eight qubits.
The developed compilation method is given the structure of the native and linear graph states and the accurate timing information to generate a quantum circuit that prepares the target graph state with high fidelity.
In \cref{fig graph state challenge}, we compare the result of the developed compilation method to the compilation provided by Qiskit with the highest optimization effort (optimization level three)~\cite{qiskit}.

We quantify the fidelity of a graph state preparation on the near-term IBM quantum computer \texttt{ibmq\_ehningen} by successively measuring the stabilizers of the graph state as described in \cref{sec graph_states}.
Each experiment was repeated sixteen times---with each compilation option equally interspersed over the experiments---to yield an accurate fidelity measurement on near-term quantum computers that inevitably exhibit large error dynamics \cite{mapomatic, not-all-qubits-are-equal, here}.
As an accurate fidelity quantification incurs a large number of quantum circuit executions in general, we decided to omit detailed results of the objective function defined in \cref{eq crosstalk}.
The experiments visualized in this section required over twelve hours of computation time on the \texttt{ibmq\_ehningen} quantum computer.

Qubit measurement errors were mitigated by the method introduced in \cite{bravyi_readout_error_mitigation}.
The solver runtime was negligible on the investigated graph state sizes and required four seconds on average for a 21-qubit linear graph state on \texttt{ibmq\_ehningen} with the objective function defined in \cref{eq runtime}.
In the remainder of the result section, the term 'SMT-Runtime' describes the combination of the Hadamard cancellation objective function in \cref{eq cancellation} and the circuit runtime objective function in \cref{eq runtime}.
The term 'SMT-Decoherence' describes the qubit decoherence objective function in \cref{eq decoherence}.

\subsection{Fidelity of the Seven-Qubit Native Graph State}

\begin{figure}[t]
    \centering
    \includegraphics[width=1.0\linewidth, trim = 0.25cm 0.35cm 0.9cm 0.8cm, clip]{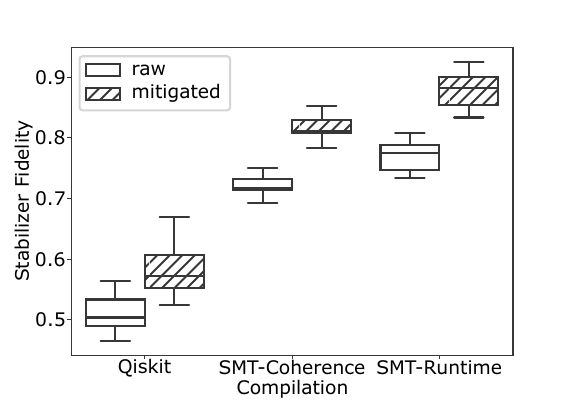}    
    \caption{Fidelity of the seven-qubit native graph state preparation after compilation with Qiskit and our method (SMT-Runtime and SMT-Decoherence). 
    The fidelity is reported with qubit-measurement error mitigation (mitigated) or without (raw) for sixteen repetitions of compilation and execution on \texttt{ibmq\_ehningen}. The figure shows box plots where the central line of the box represents the median, while each half of the box represents one quartile of the data. The whiskers show the last measured results within a distance of $1.5$ times the interquartile range.
    }
    \label{fig graph state challenge}
\end{figure}

\cref{fig graph state challenge} shows the fidelities estimated through measuring stabilizers of the seven-qubit native graph state (see \cref{fig graph state}) after compiling a graph state preparation quantum circuit using Qiskit, using SMT-Runtime and using SMT-Decoherence.
The fidelities are estimated with and without error mitigation, respectively.
We obtain fidelities that are largest when we use our compilation method together with SMT-Runtime or SMT-decoherence.
The compilation by Qiskit leads to a smaller average fidelity
even when qubit measurement error mitigation was enabled.

For all compilation methods evaluated in this work, the qubit measurement error mitigation had a significant impact on the measured fidelity.
Interestingly, the impact of the qubit measurement mitigation varies strongly for the different compilation methods.
The Qiskit graph state preparation circuit yields a median fidelity of 0.51 without qubit measurement mitigation that is improved to 0.58 with mitigation. For SMT-Runtime, we measured a fidelity of 0.77 without mitigation that is improved to 0.87 with mitigation.

Overall, the highest fidelity with measurement error mitigation was 0.93 for SMT-Runtime, while the lowest fidelity of 0.46 was observed for the Qiskit compilation without measurement mitigation.
Specifically, the SMT-Runtime compilation yielded a maximal fidelity of 0.93 and 0.88 on average, and the Qiskit compilation reached an average fidelity of 0.58 and a maximum fidelity of 0.67.
Thus, the preparation error, i.e., $1 - \text{fidelity}$, of graph state preparation, is at most reduced by a factor of 4.71x and 3.5x on average by the developed method compared to the state-of-the-art Qiskit compilation.

\subsection{Fidelity of Linear Graph States}

\begin{figure}[b]
    \centering
    \includegraphics[width=1.0\linewidth, trim = 0.25cm 0.0cm 0.9cm 0.8cm, clip]{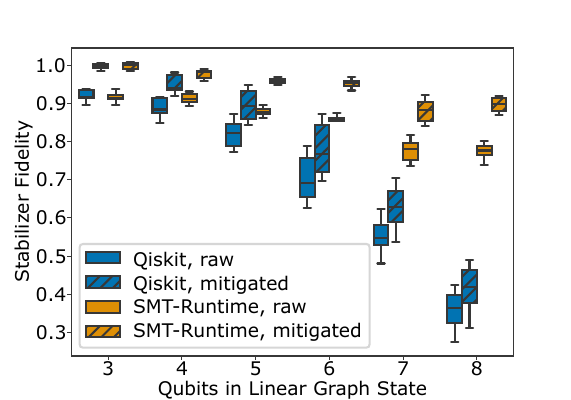}       
    
    \caption{Fidelity of linear graph states with three to eight qubits after the state preparation was compiled with Qiskit and this work (SMT-Runtime) with and without qubit measurement error mitigation for sixteen repetitions of compilation and execution on \texttt{ibmq\_ehningen}.}
    \label{fig graph state scale}
\end{figure}

\cref{fig graph state scale} shows the fidelity of linear graph states with three to eight qubits after compiling the graph state preparation quantum circuit using Qiskit and SMT-Runtime.
For three-qubit linear graph states, the fidelity yielded by the Qiskit compilation coincides with SMT-runtime as there are no additional degrees of freedom that could be exploited.
The difference in fidelity becomes more evident with larger linear graph states until it reaches a maximum at linear graph states with eight qubits. Here, the Qiskit preparation circuits show a fidelity of 0.42 on average compared to 0.91 on average, yielding a reduction in error by 6.4x on average.

Furthermore, the Qiskit graph state preparation circuits generally exhibit a larger range in fidelities compared to our method.
For eight-qubit linear graph states, the maximum fidelity is 58\% larger than the minimum fidelity achieved by the Qiskit compilation over the sixteen conducted experiments.
In SMT-Runtime, the spread between minimum and maximum fidelity is only 22\% difference.
Thus, using the developed method, higher reproducibility and consistent accuracy when preparing graph states can be expected.

The presented results align well with other results from the literature~\cite{Pathumsoot2020,deJong2024}. While in Ref.~\cite{Pathumsoot2020} quickly decaying graph-state fidelities for non-optimized circuits have been reported, similar values for circuits optimized by hand have been obtained in Ref.~\cite{deJong2024}.

\section{Conclusion\label{sec concl}}

In this work, we have presented a novel method 
for the optimized compilation of quantum circuits for the preparation of graph states on gate-based quantum computers. We have compared it to the solution provided by Qiskit. 
Our quantum circuit compilation method is based on a formal model that constructs an optimal graph state circuit by considering the specific physical architecture, accurate quantum gate timing information in the target quantum computer, gate cancellations, and gate commutations.

We have evaluated our method by producing graph states with different numbers of qubits on the IBM quantum computer \texttt{ibmq\_ehningen}, and assessed the quality of the state preparation by performing stabilizer measurements and determining the fidelity.
The presented results demonstrate an advantage of our method compared to the Qiskit solution.
Our method reduces the error when preparing a seven-qubit graph state by 3.5x on average.
For a linear eight-qubit graph state, the error is reduced by 6.4x on average.
Furthermore, it reduces the span over which the fidelities are spread over multiple experiments from 58\% to 22\%, leading to higher reproducibility and more consistent graph state preparations.

Since our method is not restricted to the generation of native graph states, one of the next steps will be to investigate the generation of other more complex quantum states. For example, it would be interesting to adapt our method to the preparation of Dicke states, which are costly but at the same time important, e.g., for the quantum alternating operator ansatz~\cite{Baertschi2019,Baertschi2022}. Furthermore, the efficient preparation of large high-fidelity GHZ states is of interest, e.g., in the context of multipartite entanglement testing quantum communication algorithms~\cite{Joy2019,Pathumsoot2020}, or benchmarking~\cite{Mooney2021GHZ,Mackeprang2023}.

Another venue for future research is the incorporation of more comprehensive noise models into the formal model used for the optimization of graph state preparations.
However, this would require more thorough noise characterization than was available publicly on \texttt{ibmq\_ehningen} at the time of experiments. 

Finally, let us note that although our compilation method has been evaluated on IBM quantum computers, it can readily be adapted to other platforms.

This work builds on one of the winning contributions to the 2020 IBM Quantum Open Science Prize~\cite{Challenge}.

\section*{Acknowledgements}
We thank Jelena Mackeprang, who participated in the original challenge, for fruitful discussions and useful suggestions. We acknowledge support from the Carl Zeiss Foundation, the Center for Integrated Quantum Science and Technology (IQ$^\text{ST}$), the Federal Ministry of Education and Research (BMBF, projects SiSiQ and PhotonQ), and the Competence Center Quantum Computing Baden-W\"urttemberg (funded by the Ministerium f\"ur Wirtschaft, Arbeit und Tourismus Baden-W\"urttemberg, project QORA). D.B. was
partially supported by the JST Moonshot R\&D program under Grant JPMJMS226C. We acknowledge the use of IBM Quantum services for this work. The views expressed are those of the authors, and do not reflect the official policy or position of IBM or the IBM Quantum team.

\bibliography{Literature}{}

\end{document}